%% file: main.tex
\documentclass[sigconf, natbib=true,screen=true, review=false]{acmart}

\acmSubmissionID{788}

\input{Preamble/packages}
\input{Preamble/authors}
\input{Preamble/meta}

\input{Preamble/definitions}

\begin{document}

\title[Model Editing for New Document Integration in Generative Information Retrieval]{Model Editing for New Document Integration in \\Generative Information Retrieval}

\begin{abstract}
Generative retrieval (GR) reformulates the Information Retrieval (IR) task as the generation of document identifiers (docIDs).
Despite its promise, existing GR models exhibit poor generalization to newly added documents, often failing to generate the correct docIDs. 
While incremental training offers a straightforward remedy,  it is computationally expensive, resource-intensive, and prone to catastrophic forgetting, thereby limiting the scalability and practicality of GR.

In this paper, we identify the core bottleneck as the decoder’s ability to map hidden states to the correct docIDs of newly added documents. 
Model editing, which enables targeted parameter modifications for docID mapping, represents a promising solution. 
However, applying model editing to current GR models is not trivial, which is severely hindered by indistinguishable edit vectors across queries, due to the high overlap of shared docIDs in retrieval results.
To address this, we propose \textbf{DOME} (docID-oriented model editing), a novel method that effectively and efficiently adapts GR models to unseen documents.
DOME comprises three stages:
\begin{enumerate*}[label=(\arabic*), leftmargin=*]
    \item identification of critical layers,
    \item optimization of edit vectors, and
    \item construction and application of updates.
\end{enumerate*}
At its core, DOME employs a \emph{hybrid-label adaptive training} strategy that learns discriminative edit vectors by combining soft labels, which preserve query-specific semantics for distinguishable updates, with hard labels that enforce precise mapping modifications.
Experiments on widely used benchmarks, including NQ and MS MARCO, show that our method significantly improves retrieval performance on new documents while maintaining effectiveness on the original collection. 
Moreover, DOME achieves this with only about 60\% of the training time required by incremental training, considerably reducing computational cost and enabling efficient, frequent model updates.\footnote{Our code is available at \url{https://github.com/zhangzhen-research/DOME}}
\end{abstract}

\maketitle

\acresetall

\input{Sections/01-intro}

\input{Sections/06-relatedwork}

\input{Sections/02-prel}
\input{Sections/03-method}

\input{Sections/04-experiment}

\input{Sections/05-results}

\input{Sections/07-conclusion}

\begin{acks}

    This research was (partially) supported by the Natural Science Foundation of China (62272274, 62202271, 62472261, 62372275, 62522210, T2293773), the National Key R\&D Program of China with grant No. 2024YFC3307300 and No. 2022YFC3303004, the Provincial Key R\&D Program of Shandong Province with grant No. 2024CXGC010108, the Natural Science Foundation of Shandong Province with grant No. ZR2024QF203, the Technology Innovation Guidance Program of Shandong Province with grant No. YDZX2024088,
    the Dutch Research Council (NWO), under project numbers 024.004.022, NWA.1389.20.\-183, and KICH3.LTP.20.006, and the European Union under grant agreements No. 101070212 (FINDHR) and No. 101201510 (UNITE).
    
    All content represents the opinion of the authors, which is not necessarily shared or endorsed by their respective employers and/or sponsors.
\end{acks}

\acresetall

\clearpage
\bibliographystyle{ACM-Reference-Format}
\balance
\bibliography{references}
\clearpage

\appendix
\input{Sections/08-appendix}

\end{document}

%% file: Preamble/packages.tex
\usepackage{booktabs} 
\usepackage{algorithm, algpseudocode}
\usepackage{amsmath}
\usepackage{graphics}
\usepackage{epsfig}
\usepackage{graphicx}
\usepackage{xcolor}
\usepackage[skip=1pt]{caption}
\usepackage{subfigure}
\usepackage{balance}
\usepackage{multirow}
\usepackage{mathrsfs}
\usepackage{acronym}
\usepackage{placeins}
\usepackage{xcolor}
\usepackage[inline]{enumitem}

\usepackage{newfloat}
\DeclareFloatingEnvironment[fileext=lop,listname={List of prompts},name=Prompt, placement=h]{prompt}
\usepackage{adjustbox}
\usepackage{graphicx}
\usepackage{wrapfig}
\usepackage{booktabs}

\usepackage[table]{xcolor}
\usepackage{xspace}
\usepackage{bm}

%% file: Preamble/authors.tex
\author{Zhen Zhang}
\orcid{0009-0001-0290-5386}
\affiliation{%
  \institution{Shandong University}
  \city{Qingdao}
  \country{China}
}
\email{zhen.zhang.sdu@gmail.com}

\author{Zihan Wang}
\authornote{Corresponding author.}
\orcid{0000-0003-0493-2668}
\affiliation{%
  \institution{University of Amsterdam}
  \city{Amsterdam}
  \country{The Netherlands}
}
\email{zhw.cypher@gmail.com}

\author{Xinyu Ma}
\orcid{0000-0002-5511-9370}
\affiliation{%
  \institution{Baidu Inc.}
  \city{Beijing}
  \country{China}
}
\email{xinyuma2016@gmail.com}

\author{Shuaiqiang Wang}
\orcid{0000-0002-9212-1947}
\affiliation{%
  \institution{Baidu Inc.}
  \city{Beijing}
  \country{China}
}
\email{shqiang.wang@gmail.com}

\author{Dawei Yin}
\orcid{0000-0002-0684-6205}
\affiliation{%
  \institution{Baidu Inc.}
  \city{Beijing}
  \country{China}
}
\email{yindawei@acm.org}

\author{Xin Xin}
\orcid{0000-0003-4703-7356}
\affiliation{%
  \institution{Shandong University}
  \city{Qingdao}
  \country{China}
}
\email{xinxin@sdu.edu.cn}

\author{Pengjie Ren}
\orcid{0000-0003-2964-6422}
\affiliation{%
  \institution{Shandong University}
  \city{Qingdao}
  \country{China}
}
\email{renpengjie@sdu.edu.cn}

\author{Maarten de Rijke}
\orcid{0000-0002-1086-0202}
\affiliation{%
  \institution{University of Amsterdam}
  \city{Amsterdam}
  \country{The Netherlands}
}
\email{m.derijke@uva.nl}

\author{Zhaochun Ren}
\authornotemark[1]
\orcid{0000-0002-9076-6565}
\affiliation{%
  \institution{Leiden University}
  \city{Leiden}
  \country{The Netherlands}
}
\email{z.ren@liacs.leidenuniv.nl}

\renewcommand{\shortauthors}{Zhen Zhang et al.}

%% file: Preamble/meta.tex
\copyrightyear{2026}
\acmYear{2026}
\setcopyright{cc}
\setcctype{by}
\acmConference[WWW '26]{Proceedings of the ACM Web Conference 2026}{April 13--17, 2026}{Dubai, United Arab Emirates}
\acmBooktitle{Proceedings of the ACM Web Conference 2026 (WWW '26), April 13--17, 2026, Dubai, United Arab Emirates}
\acmPrice{}
\acmDOI{10.1145/3774904.3792178}
\acmISBN{979-8-4007-2307-0/2026/04}

\ccsdesc[500]{Information systems~Retrieval models and ranking}

\keywords{Model editing, Generative retrieval}

%% file: Preamble/definitions.tex
\newcommand{\headernodot}[1]{\vspace*{1mm}\noindent\textbf{#1}}
\newcommand{\header}[1]{\headernodot{#1.}}

%% file: Sections/01-intro.tex
\section{Introduction}
Generative retrieval (GR) marks a paradigm shift in information retrieval. Instead of searching a pre-built index, GR models directly generate relevant document identifiers (docIDs) in response to a query~\citep{tay2022transformer, wang2022neural, bevilacqua2022autoregressive, tang2023semantic, zhou2024roger}. 
GR is gaining increasing attention in the IR community as it enables end-to-end training and shows excellent retrieval performance ~\citep{de2020autoregressive, zeng2024planning, zhang2024generative, tang2024generative}.
However, a critical limitation hinders the practical deployment of GR: poor generalization to new documents, with models often failing to retrieve items added after the initial training phase~\citep{mehta2022dsi++, kim2023exploring, zhang2025replication, zhang2025does, wu2025constrained}.
A prevalent strategy to address this problem is incremental training, in which the GR model is retrained on new additions when documents are added to the collection~\citep{mehta2022dsi++, guo2024corpusbrain++}. 
While this enables the model to learn docIDs for new entries, it is computationally expensive, requiring substantial data and resources. 
Moreover, it is prone to catastrophic forgetting~\citep{mehta2022dsi++, kim2023exploring, liu2023robustness}, where adapting to new documents diminishes performance on the initial set. 
These drawbacks significantly limit the scalability and practicality of current GR models.

\begin{figure}[tbp]
\centering
\includegraphics[width=\linewidth]{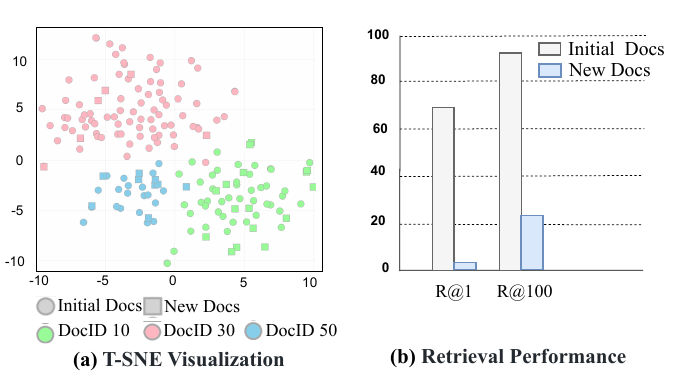}
\caption{
Behavioral analysis of the initial and new documents over the NQ dataset~\citep{kwiatkowski2019natural} using a hierarchical k-means-based docID GR model~\citep{tay2022transformer}.
The model is trained on the initial document set, and docIDs for new documents are assigned using the same k-means clustering procedure.
(a) T-SNE~\citep{maaten2008visualizing} visualization of intermediate representations, where different colors denote different docID prefixes (i.e., 10, 30, 50), with each prefix corresponding to a specific document type. Squares represent initial documents, and circles represent new documents.
(b) Retrieval performance (Recall@1 and Recall@100) on the initial and new document sets.}
\label{fig:similar}
\end{figure}

In this work, we first conduct a detailed analysis to identify the root cause of the poor performance of GR models on newly added documents.
Our findings reveal that the key bottleneck lies in the decoder's capacity to map hidden states to the correct docIDs of these documents. 
As shown in Figure~\ref{fig:similar}(a), the encoder is able to position the newly added documents close to semantically related ones from the initial training set in the representation space. 
This indicates that the model has successfully captured the semantic content of the new documents. 
In contrast, Figure~\ref{fig:similar}(b) shows a marked degradation in retrieval accuracy for the newly added documents.
This discrepancy indicates that the mapping between semantic representations and specific docIDs is established only during initial training, but fails to generalize to new documents.
Consequently, full retraining is unnecessary: The key is to update this mapping for new documents while maintaining retrieval effectiveness on the initial set.

This insight naturally motivates \textbf{model editing} as a promising solution, as it enables targeted parameter modifications for docID mapping while avoiding the high computational cost and catastrophic forgetting of incremental training~\citep{mitchell2022memory, dong2022calibrating}.
However, applying model editing to GR is undermined by a fundamental problem: the edit vectors for different queries lack discriminability. 
In typical editing of subject–relation–object triplets (denoted as $<s, r, o>$), the target objects are often unique.
In contrast, GR generates docIDs autoregressively as a sequence of discrete tokens. 
At each step, the model predicts the next token (target docID) conditioned on the query and the previously generated tokens (prefix docID). 
The edit requests in GR take the form of <query, prefix docID, target docID> mappings.
A key challenge arises from the limited docID vocabulary, which inevitably causes many different queries and docID prefixes to be mapped to the same target docID.
Consequently, the edit vectors lose their discriminability for distinct GR contexts.
This prevents the model from learning the precise mapping for each query that severely degrades editing accuracy.

To tackle the above challenges, we introduce \textbf{DOME} (docID-oriented model editing), a GR-specific editing method designed to efficiently adapt models to new documents without full retraining. 
DOME operates in three stages:
First, the decoder FFN modules that encode docID mapping knowledge are identified through average patching analysis~\citep{reusch2025reverse}, ensuring that updates are confined to the most relevant parameters.
Second, edit vectors for new docIDs are optimized using a \emph{hybrid-label adaptive training} strategy that combines soft labels from the model’s output distribution, which preserve query-specific semantic diversity, with hard ground-truth labels that enforce precise generation of the correct docID tokens.
During training, the emphasis gradually moves from soft labels to hard labels, encouraging discriminative updates at the start and ensuring precise docID generation by the end.
Finally, the optimized edit vectors are assembled into constrained updates and applied to the selected FFN modules, allowing the model to incorporate mappings for new documents while maintaining retrieval performance on the original collection.
Extensive experiments on two standard benchmarks, Natural Questions (NQ) and MS-MARCO, demonstrate that our method significantly improves retrieval accuracy for new documents while exhibiting strong resilience to catastrophic forgetting. 
Moreover, DOME significantly reduces the adaptation time compared with conventional incremental training, enabling much faster deployment in real-world retrieval scenarios.

Our contributions are summarized as follows:
\begin{itemize}[leftmargin=*]
\item We identify that the core bottleneck in adapting GR to new documents lies in the model's inability to generate newly assigned docIDs, even when it correctly understands their content. 
\item We propose a hybrid-label adaptive training strategy to address the issue of insufficient discriminability in edit vectors that arises when applying model editing to GR.
\item We develop DOME, a specialized and efficient model editing framework for GR, achieving strong retrieval performance on new documents while substantially mitigating catastrophic forgetting and reducing adaptation costs.
\item Extensive experiments on NQ and MS-MARCO show that \textbf{DOME} consistently outperforms incremental training, while reducing adaptation time by over 40\%. 
\end{itemize}

%% file: Sections/06-relatedwork.tex
\section{Related Work}

\subsection{Generative retrieval}

Generative retrieval (GR) formulates document search as sequence generation, where the model directly outputs a \emph{docID} given a query, replacing index-based lookup.  
DocIDs can be \emph{textual} identifiers such as titles or pseudo queries~\citep{zhou2022ultron, bevilacqua2022autoregressive, li2023multiview}, or \emph{numeric} codes composed of discrete tokens~\citep{sun2023learning, yang2023auto, zhou2023dynamicretriever, zhao2025diffugr, mekonnen2025lightweight}.  
Textual IDs are flexible but incur costly string matching~\citep{wang2024generative, wang2023novo}.  
Numeric IDs are typically derived via clustering or quantization, e.g., $k$-means~\citep{tay2022transformer}, PQ~\citep{zhou2022ultron}, RQ~\citep{Zeng2023ScalableAE}, or atomic single-number IDs~\citep{kishore2023incdsi}.  
We focus on structured numeric docIDs (e.g., PQ or RQ), which offer semantic grouping and scalability over atomic IDs.

A key challenge in GR is generalization to unseen documents~\citep{kim2023exploring}.  
DSI++~\citep{mehta2022dsi++} extends DSI to structured numeric docIDs and improves unseen-document generalization via mixed training and flattened loss.  
incDSI~\citep{kishore2023incdsi} enables real-time insertion through constrained parameter updates but supports only atomic IDs and requires codebook expansion, limiting applicability to structured docIDs.

\subsection{Model editing}
Model editing updates internal knowledge through small targeted parameter changes without full retraining~\citep{mazzia2024survey}.  
Meta-learning approaches such as KE~\citep{de2021editing} and MEND~\citep{mitchell2022memory} use hypernetworks to generate gradient-based edits, enabling fast updates but affecting many parameters.  
Locate--then--edit methods identify knowledge-bearing components and update only selected parameters~\citep{santurkar2021editing, xu2022language}.  
ROME~\citep{meng2022locating} locates FFN neurons encoding specific facts and applies closed-form updates, while MEMIT~\citep{meng2022mass} supports batch edits.  
AlphaEdit~\citep{fang2024alphaedit} further introduces orthogonal projections to reduce forgetting in lifelong editing.

In this work, we propose \textbf{DOME}, a GR adaptation framework that leverages model editing to incorporate new documents after initial training.  
Unlike incremental retraining~\citep{mehta2022dsi++, kim2023exploring,DBLP:conf/emnlp/LyuYWSYRCRR24,DBLP:conf/iclr/LyuY0YRRR25}, DOME avoids expensive updates and mitigates forgetting through selective parameter edits.  
In contrast to existing editing methods~\citep{meng2022locating, meng2022mass, fang2024alphaedit}, which struggle with GR due to indistinguishable edit vectors, DOME introduces a hybrid-label adaptive training strategy that combines soft-label semantic diversity with hard-label precision, gradually shifting to pure hard labels for exact docID mapping.  
This GR-specific design enables efficient integration of new identifiers while preserving retrieval performance on both original and newly added documents.

%% file: Sections/02-prel.tex
\section{Preliminaries}  
\label{sec:task_formulation}
\subsection{Generative retrieval}
Generative Retrieval (GR) formulates document retrieval as a sequence generation task.
Given a query $q$, a GR model directly generates the document identifier (docID) of the most relevant document~\citep{de2020autoregressive}.
The docID is represented as a sequence of discrete tokens $\bm{y} = (y_1, \dots, y_T)$.
These tokens are often assigned via structured schemes such as hierarchical $k$-means clustering~\citep{ahmed2020k}.

The decoder generates the docID sequence $\bm{y}$ auto-regressively, predicting each token $y_t$ conditioned on the query and the docID prefix $y_{<t}$ as follows:
\begin{equation}
    p(\bm{y}|q) = \prod_{t=1}^{T} p(y_t | q, y_{<t}).
\end{equation}
The training objective is to maximize this conditional likelihood, which corresponds to minimizing the negative log-likelihood loss for a query-document pair $(q, d)$:
\begin{equation}
    \mathcal{L}_{\text{GR}} = -\sum_{t=1}^{T} \log p(y_t | q, y_{<t}; \theta),
    \label{eq:gr-loss}
\end{equation}
where $\theta$ represents the GR model parameters~\citep{sutskever2014sequence}.

\header{Adapting to new documents} 
Our goal is to adapt a pre-trained GR model, originally trained on a corpus $\mathcal{C}_{\text{init}}$, to effectively retrieve documents from a newly added corpus $\mathcal{C}_{\text{new}}$~\citep{mehta2022dsi++}. 
The model’s parameters allow it to correctly generate docIDs for documents in $\mathcal{C}_{\text{init}}$, but it often fails to produce the correct docID sequence for a query $q$ whose relevant document lies in $\mathcal{C}_{\text{new}}$. 
Instead of relying on incremental training, we formulate this adaptation as a model editing problem, where the model is updated with new docID mappings for $\mathcal{C}_{\text{new}}$ while preserving its performance on $\mathcal{C}_{\text{init}}$.

\subsection{Model editing}
\label{subsec:model editing}
Model editing aimsg to efficiently update specific knowledge within a language model without full retraining, while preserving its original knowledge and performance~\citep{jiang2025anyedit, khandelwal2024cross}.
In typical settings, stored knowledge is represented as subject–relation–object triplets $< s, r, o>$~\citep{cheng2024editing,DBLP:conf/cikm/0002ZHCRRR23,DBLP:conf/ijcai/WangRHZH19,DBLP:conf/cikm/HeWZTR20,DBLP:conf/emnlp/WangZCRRR23}, and an editing request specifies a behavioral modification such that the model outputs the updated object $o$ for a given $(s, r)$ pair~\citep{zheng2023can}.

In transformer architectures, a feed-forward network (FFN) module can be interpreted as a \emph{key–value memory}~\citep{geva2020transformer}: the \emph{key vector} $k$ encodes the subject–relation pair $(s, r)$, while the \emph{value vector} $v$ encodes the object $o$. 
Specifically, given a hidden input state $h$, 
the input projection $W_{\mathrm{in}}$ followed by a non-linear activation $\sigma$ produces a \emph{key} vector:
\begin{equation}
    k = \sigma(W_{\mathrm{in}} h) \in \mathbb{R}^{d_0},
\end{equation}
where $d_0$ denotes the dimension of the FFN’s intermediate layer.  
The output projection $W_{\mathrm{out}} \in \mathbb{R}^{d_1\times d_0}$ then maps $k$ to a \emph{value} vector:
\begin{equation}
    m = W_{\mathrm{out}} k \in \mathbb{R}^{d_1},
\end{equation}
where $d_1$ denotes the dimension of the FFN’s output layer.

Model editing modifies the output projection $W_{\mathrm{out}}$ by adding an update matrix $\Delta$.
Suppose we aim to perform $u$ edits to inject new knowledge. 
For each edit $<s_i, r_i, o_i>$, we obtain a new key–value pair $(k_i, v_i)$ representing the desired knowledge modifications, where each $k_i$ is computed by feeding the context $(s_i, r_i)$ into $W_{\mathrm{in}}$ followed by the non-linear activation $\sigma$, and $v_i$ is the desired value, namely the target output vector for $o_i$ that the model should produce.
Stacking all $u$ keys and values yields $K_1 = [k_1 \mid k_2 \mid \dots \mid k_u] \in \mathbb{R}^{d_0 \times u}$ and $V_1 = [v_1 \mid v_2 \mid \dots \mid v_u] \in \mathbb{R}^{d_1 \times u}$, where $K_1$ denotes the keys and $V_1$ their desired values~\citep{vaswani2017attention}.

To obtain $v_i$, we first compute the model’s original output
\begin{equation}
    m_i = W_{\mathrm{out}} k_i,
\end{equation}
and then determine a minimal editing vector $\delta_i$ that shifts the prediction toward the correct object $o_i$:
\begin{equation}
\label{eq:model-edit-objective-new}
    \delta_i = \arg\min_{\delta} -\log p_{\theta,\, m_i+\delta}(o_i \mid s_i, r_i),
\end{equation}
where $p_{\theta, m_i+\delta}$ denotes the model's output distribution when the FFN output is replaced with $m_i+\delta$. The updated value is then given by $v_i = m_i + \delta_i$, and stacking all updated keys and values yields the complete $K_1$ and $V_1$.

To mitigate forgetting~\citep{dai2021knowledge}, we additionally construct a set of preserved key–value pairs $(K_0, V_0)$ representing knowledge that should remain unchanged. 
Here, $K_0$ is obtained in the same way as $K_1$ but from original knowledge, and $V_0 = W_{\mathrm{out}} K_0$ stores their original outputs. The update matrix $\Delta$ is then computed to align $(K_1, V_1)$ while minimally disturbing $(K_0, V_0)$:
\begin{equation}
    \Delta = \arg\min_{\tilde{\Delta}} \;\| (W_{\mathrm{out}} + \tilde{\Delta})K_1 - V_1 \|_F^2
    + \|\tilde{\Delta} K_0\|_F^2 ,
\end{equation}
where $\|\cdot\|_F$ denotes the Frobenius norm~\citep{bottcher2008frobenius}. This objective has the closed-form solution:
\begin{equation}
    \Delta = (V_1 - W_{\mathrm{out}} K_1)\left(K_1 K_1^\top + K_0 K_0^\top\right)^{-1} K_1^\top.
\label{eq:model-edit-solution-new}
\end{equation}
In practice, $K_0$ can be estimated from abundant auxiliary inputs in the original corpus to approximate the preserved knowledge subspace.

\header{Editing for GR} 
In generative retrieval, the editing process targets the auto-regressive generation of document identifiers~\citep{tay2022transformer}.  
Each edit request can be represented as $< q, y_{<t}, y_t >$, where $q$ is the input query, $y_{<t}$ is the generated docID prefix up to step $t{-}1$, and $y_t$ is the desired next docID token.  
In this setting, the FFN key $k$ is computed from the decoder hidden state conditioned on $(q, y_{<t})$, and the corresponding value $v$ encodes the target docID token $y_t$~\citep{vaswani2017attention}.  
By stacking multiple such keys and values into matrices $K_1 \in \mathbb{R}^{d_0 \times u}$ and $V_1 \in \mathbb{R}^{d_1 \times u}$, the editing framework can directly adjust the model’s internal key–value representations to produce the intended outputs for the given query–prefix contexts.

%% file: Sections/03-method.tex
\section{Discussion}
Before presenting our proposed method, we conduct a comprehensive analysis to identify the docID mapping bottleneck that limits the adaptation of current GR models to new documents (see \S\ref{section:3.1}).
We further point out that existing model editing methods cannot fully resolve this issue, as their effectiveness is severely hindered by indistinguishable edit vectors across queries (see \S\ref{section:3.2}).

\subsection{DocID mapping bottleneck}
\label{section:3.1}
Recent mechanistic analyses of generative retrieval (GR)~\citep{reusch2025reverse,lee2024these} show that the encoder captures the semantic content of inputs, while the decoder generates the corresponding docIDs.  
The decoding process can be conceptually divided into three stages: (1) a priming stage that initializes the task representation, (2) a bridging stage where cross-attention extracts semantics from the encoder outputs, and (3) an interaction stage where FFN modules transform semantic features into final docID tokens~\citep{reusch2025reverse}.

\input{Tables/03-docid_accuracy}

\noindent%
As Figure~\ref{fig:similar}(a) shows, although the encoder produces meaningful representations for new documents, the decoder fails to generate their correct docIDs. 
To analyze this failure, we examine docID-level prediction accuracy.  
As shown in Table~\ref{tab:doc_accuracy}, the decoder achieves 42.3\% accuracy on individual docID tokens but only 17.3\% on complete docID sequences.  
This gap indicates that the model learns an approximate rather than an exact mapping from semantics to docIDs. It can predict partial tokens but not the precise sequence required for retrieval.  
Hence, the core challenge in adapting GR models to new documents lies in their inability to adjust this learned docID mapping.  
Therefore, instead of retraining the entire model, it is more effective to selectively update the decoder parameters responsible for docID mapping, thereby overcoming this bottleneck and improving retrieval for newly added documents.

\begin{figure}[h]
    \centering
    \includegraphics[width=\linewidth]{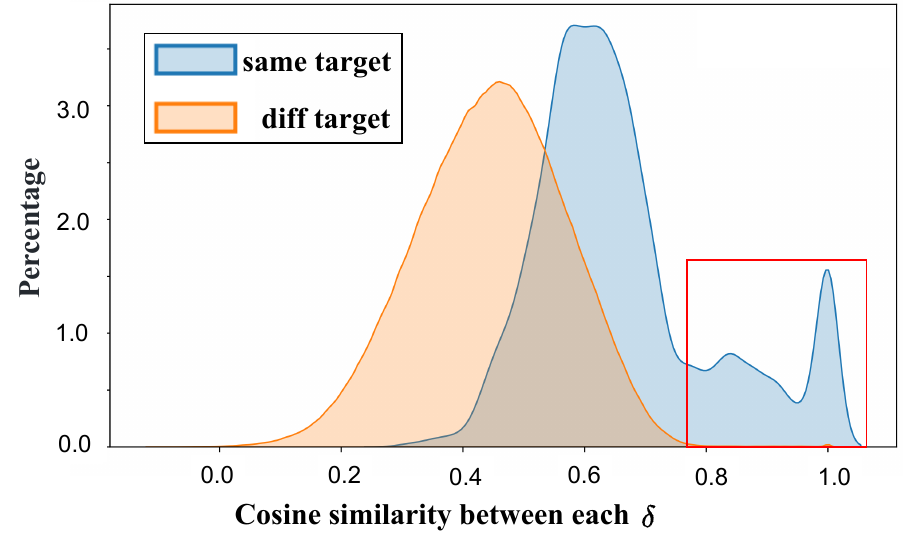}
    \caption{Percentage of pairwise cosine similarity of edit vector $\delta$ across GR edit requests applied to the GR method~\citep{tay2022transformer} on NQ~\citep{kwiatkowski2019natural}. Yellow denotes pairs with different target docID, blue denotes pairs with the same target docID.}
    \label{fig:delta_similar}
\end{figure}

\begin{figure*}[ht]
    \centering
    \includegraphics[width=0.93\textwidth]{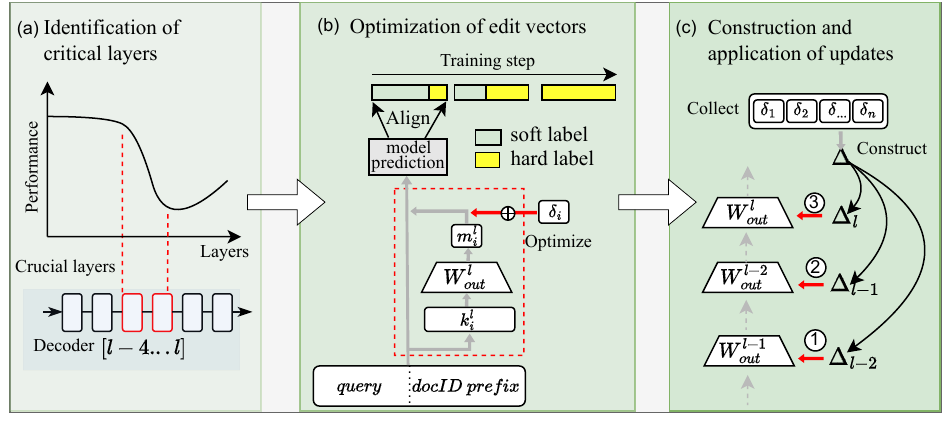}
    \caption{
An overview of the proposed DOME framework. 
(a) A patching technique is used to diagnose the model and locate the critical layers responsible for docID mapping. 
(b) Our hybrid-label adaptive training strategy is employed to compute diverse and effective edit vectors ($\delta$) for the new mappings.  
For clarity, the diagram retains only the crucial $W_{\mathrm{out}}$ component from the FFN layer.
(c) These edit vectors are then used to construct an update matrix ($\Delta$), which is applied to the crucial layer parameters ($W_{\text{out}}$) to inject the new docID mapping.}
    \label{fig:method}
\end{figure*}

\subsection{Indistinguishable edit vectors in GR}
\label{section:3.2}
As established in \S\ref{section:3.1}, GR models struggle to generate exact docIDs for new documents due to misaligned mappings between encoded semantics and discrete docID tokens.
Model editing provides a promising solution by enabling targeted parameter updates without full retraining, but its effectiveness in GR is limited by indistinguishable edit vectors across queries.
Specifically, as mentioned in \S\ref{subsec:model editing}, each GR editing request can be represented as a triple $\langle q, y_{<t}, y_t\rangle$, where $(q, y_{<t})$ is the input context and $y_t$ is the target docID token.  
Because docID vocabularies are limited, many distinct contexts $(q, y_{<t})$ map to the same target token $y_t$.  
When optimized under a one-hot label objective (Eq.~\ref{eq:model-edit-objective-new}), these requests share identical supervision signals, leading to nearly identical edit vectors $\delta_i$ even for semantically distinct queries.  
This lack of discriminability prevents the model from learning context-specific query–docID mappings.

To investigate this, we construct GR editing requests for new documents, compute their edit vectors $\delta$ from the targeted FFN layers, and measure pairwise cosine similarity~\citep{xia2015learning}. 
As shown in Figure~\ref{fig:delta_similar}, requests sharing the same target token exhibit high similarity between their edit vectors, confirming that one-hot hard-label optimization produces nearly indistinguishable updates, making it difficult for the model to learn precise docID mappings.

\section{DOME: DocID-Oriented Model Editing for Generative Retrieval}
\label{rec:dome}



\textbf{DOME} is a GR‑specific model editing framework designed to inject new docID mappings into a pre‑trained GR model while preserving its original retrieval ability.
As shown in Figure~\ref{fig:method}, DOME consists of three stages:
(a) Identification of critical layers: locating the decoder’s critical FFN layers responsible for docID mapping via average patching technology;
(b) Optimization of edit vectors: computing diverse and effective edit vectors using a \emph{hybrid‑label adaptive training} strategy, thereby mitigating the indistinguishable edit vectors problem.
(c) Construction and application of updates: assembling these edit vectors into a constrained update matrix and sequentially applying it to the targeted layers to inject new mappings while retaining existing knowledge.

\subsection{Identification of critical layers}
\label{section:3.3}
Figure~\ref{fig:method}(a) illustrates the first stage of our workflow, which aims to locate the crucial layers that store docID‑mapping knowledge.
We adopt an \emph{average patching} strategy to identify these critical layers~\citep{reusch2025reverse}.
Specifically, we first collect the output representations of all modules in the decoder’s intermediate layers over a document set.
Then, for another set of documents, we replace each layer’s original output with the average representation computed from the training data.
By measuring the retrieval performance drop after each replacement, we determine which five layers have the greatest impact when patched.
We regard the five layers from \(l-4\) to \(l\) as critical for docID mapping, as altering their outputs significantly degrades retrieval accuracy.
Subsequent model‑editing operations are performed exclusively on these identified layers to ensure targeted and efficient adaptation.
Detailed analyses are provided in Appendix~\ref{appdix:patching}.

\subsection{Optimization of edit vectors}
Figure~\ref{fig:method}(b) illustrates the second stage of DOME, which computes the edit vectors for injecting new docID mappings into the critical decoder layers identified in Section~\ref{section:3.3}.  
We construct a series of GR editing requests from the new documents.  
For each new document, we generate pseudo‑queries whose relevant target is this document, and pair them with its assigned docID sequence $(y_{1}, \dots, y_{T})$~\citep{nogueira2019doc2query}.  
Following the request format defined in Section~\ref{subsec:model editing}, each request is represented as $< query, docID\ prefix, target\ docID >$.  

At the core of DOME is a \textbf{hybrid‑label adaptive training} strategy, designed to address the low discriminability of edit vectors in GR identified in \S\ref{section:3.2}.  
Our strategy directly tackles the low discriminability of $\delta$ by replacing the one‑hot hard label with a hybrid label that mixes the model's original output distribution (soft label) with the ground‑truth token (hard label)~\citep{muller2019does}:
\begin{equation}
p_{\mathrm{target}}^{(s)}(v \mid q, y_{<t}) = (1 - \lambda_s) \, p_{\text{orig}}(v \mid q, y_{<t}) + \lambda_s \, \mathbf{1}[v = y_t],
\label{eq:hybrid-label}
\end{equation}
where $v$ is the docID vocabulary token, $p_{\text{orig}}$ is the model's original output distribution, $\mathbf{1}[v = y_t]$ is the one‑hot distribution of the correct token, and $\lambda_s \in [0,1]$ gradually increases from a small initial value to $1.0$ as training proceeds.  
The soft component preserves query‑specific diversity across edit vectors, while the hard component enforces accurate prediction.

For each token‑level editing request, we compute the edit vector \(\delta_i\) at the final critical layer \(l\).  
The hybrid label is then used to optimize the edit vector $\delta_i$ by modifying the optimization in Eq.~\ref{eq:model-edit-objective-new} to:
\begin{equation}
\small
\label{eq:model-edit-objective-new}
\mbox{}\hspace*{-6pt}
    \delta_i \!=\! \arg\min_{\delta_i} \!\left( - \!\sum_{v \in \mathcal{V}} p_{\mathrm{target}}^{(s)}(v \mid q_i, y_{<t,i}) \log p_{\theta, m_i+\delta_i}(v \!\mid\! q_i, y_{<t,i}) \!\right),
    \hspace*{-6pt}\mbox{}
\end{equation}
where $p_{\theta, m_i+\delta_i}$ denotes the model's output distribution when the FFN output is patched with $m_i+\delta_i$.  
Since $p_{\text{orig}}$ varies across queries, the resulting $p_{\mathrm{target}}^{(s)}$ and optimal $\delta_i$ also differ, encouraging diverse update directions in the early stage.  
As $\lambda_s$ approaches $1.0$, training focuses increasingly on the hard label, ensuring that the learned mappings generate the exact docID tokens for new documents.

\subsection{Construction and application of updates}
Figure~\ref{fig:method}(c) illustrates the final stage of DOME, where the optimized edit vectors are integrated into the targeted decoder layers while avoiding catastrophic forgetting.  
For each identified FFN layer $l$, we collect two sets of key–value pairs:  
$K_1^{(l)}$ contains the key activations from the new GR editing requests, and $V_1^{(l)}$ stores the corresponding target values obtained by adding the optimized edit vectors $\delta$ to the FFN’s original outputs.  
To preserve existing knowledge, we also gather $K_0^{(l)}$ and $V_0^{(l)} = W_{\mathrm{out}}^{(l)}K_0^{(l)}$ from query–docID pairs in the original corpus.

Our objective is to align the output of the final identified layer $l$ with $V_1^{l}$.  
Following the constrained closed‑form solution in Eq.~\ref{eq:model-edit-solution-new}, we first compute the update matrix $\Delta$ for this final layer directly using $K_0^{l}$ and $K_1^{l}$.  

To further reduce the risk of knowledge forgetting, DOME distributes this update across multiple preceding layers, similar to strategies used in prior work.  
Specifically, for the five layers before the final target layer, the update matrix at layer $j$ is scaled proportionally to the inverse of its distance from $l_{\mathrm{f}}$:  
\begin{equation}
\small
\begin{split}
\Delta^{(j)} 
= &\frac{1}{\mathrm{dist}(j, l)}
\left( V_1^{(l)} - W_{\mathrm{out}}^{(l)} K_1^{(l)} \right) \\
&\left( K_0^{(j)} {K_0^{(j)}}^\top + K_1^{(j)} {K_1^{(j)}}^\top 
\right)^{-1}{K_1^{(j)}}^\top ,
\end{split}
\end{equation}
where $\mathrm{dist}(j, l) = |\,l - j + 1\,|$ denotes the absolute layer distance~\citep{meng2022mass}.  
These updates are applied sequentially from the earliest of the targeted layers toward the final layer $l$.  
After applying $\Delta_j$ at a given layer, we re‑run the forward pass to refresh the key activations $K_1^{(l)}$ for subsequent layers, ensuring that each update is based on up‑to‑date intermediate representations.  
This stage precisely injects the new docID mappings into the decoder, preserving established retrieval capabilities and completing DOME’s adaptation to the newly added corpus.

%% file: Tables/03-docid_accuracy.tex
\begin{table}[htbp]
\centering
\caption{Comparison of Recall@10 and accuracy rates of the GR model~\citep{tay2022transformer} for initial and added documents on NQ~\citep{kwiatkowski2019natural}.}
\label{tab:doc_accuracy}
\begin{tabular}{lcccc}
\toprule
& & \multicolumn{3}{c}{\textbf{Accuracy Rate of the i\textsuperscript{th} DocID}} \\
\cline{3-5}
 & \textbf{Recall@10} & \textbf{DocID 1} & \textbf{DocID 2} & \textbf{DocID 3}  \\
\midrule
Initial Docs & 0.862 & 0.774 & 0.772 & 0.757  \\
New Docs   & 0.173 & 0.423 & 0.385 & 0.420  \\
\bottomrule
\end{tabular}
\end{table}

%% file: Sections/04-experiment.tex
\section{Experiments}
We aim to answer the following research questions:
\begin{enumerate*}[label=(\textbf{RQ\arabic*}),leftmargin=*,nosep]
\item Does DOME effectively and efficiently adapt generative retrieval (GR) models to newly added documents?
\item Does DOME preserve retrieval performance on the original document set, thereby mitigating catastrophic forgetting?
\item Does each core component of DOME contribute to its overall performance?
\item Does DOME maintain stable performance as the number of newly added documents increases?\footnote{Due to space constraints, the detailed analysis for RQ4 is provided in Appendix~\ref{appdix:rq4}.}
\end{enumerate*}

\subsection{Experimental setup}
\noindent\textbf{Baselines.}
We compare DOME with three categories of baselines: 
\begin{enumerate*}[leftmargin=*,nosep]
    \item \textbf{No training.} This category keeps all model parameters frozen after adding new documents, relying solely on the model's inherent generalization ability. It includes the sparse retriever \textbf{BM25}~\citep{robertson2009probabilistic}, the dense retriever \textbf{DPR}~\citep{karpukhin2020dense}, and the generative retrievers \textbf{DSI}~\citep{tay2022transformer} and \textbf{Ultron}~\citep{zhou2022ultron}, all used without parameter updates.
    \item \textbf{Incremental training.} These methods update model parameters through additional training to incorporate new document knowledge. They include \textbf{From Scratch}, which trains a GR model on the combined corpora; \textbf{New-Doc FT}~\citep{mehta2022dsi++}, which fine-tunes the initial GR model using queries and pseudo-queries for new documents; and \textbf{DSI++}~\citep{mehta2022dsi++}, which fine-tunes on both initial and new queries while promoting flatter loss basins to mitigate catastrophic forgetting.
    \item \textbf{Model editing.} These methods directly inject new knowledge into targeted parameters without full retraining. We include \textbf{ROME}~\citep{meng2022locating}, \textbf{MEMIT}~\citep{meng2022mass}, and \textbf{AlphaEdit}~\citep{fang2024alphaedit}, originally designed for knowledge editing but adapted here for GR by redefining edit requests as docID mappings and limiting updates to critical FFN modules.
\end{enumerate*}
For fairness, all incremental training and model editing baselines start from the same GR model trained on the original corpus and adapt to the same set of newly added documents.

\noindent\textbf{Datasets.}
We conduct experiments on two standard benchmarks: \textbf{Natural Questions (NQ)}~\citep{kwiatkowski2019natural} and \textbf{MS-MARCO}~\citep{nguyen2016ms}. 
To simulate the addition of new documents in real-world settings, each corpus is partitioned into 90\% for training the base GR model and 10\% as the new document set. 
This setup allows us to evaluate the model's ability to adapt to new documents while retaining retrieval performance on the original collection.

\noindent\textbf{Metrics.}
Retrieval performance is evaluated using Recall@K and Mean Reciprocal Rank (MRR). 
Following~\citep{mehta2022dsi++}, we further compute the forgetting score ($F_n$) based on R@10 to quantify performance degradation on the initial corpus. 
Efficiency is measured by the update time per document (TPD), defined as the total training or editing time divided by the number of newly added documents.

\input{Tables/00-main}

\noindent\textbf{Implementation Details.}  
Experiments are conducted on four NVIDIA A100 GPUs. 
We adopt T5-large\footnote{\url{https://huggingface.co/google-t5/t5-large}} as the base GR model, jointly optimizing query to docID, document to docID, and pseudo-query to docID generation tasks. 
Pseudo-queries for new documents are generated using the docTTTTTquery model~\citep{nogueira2019doc2query}.  
For DOME, based on our layer importance analysis, editing is applied to the FFN modules in decoder layers 14 to 18. 
In the hybrid label adaptive training strategy, the mixing coefficient $\lambda_s$ increases linearly from 0.3 to 1.0 over 50 optimization steps. 
We use residual quantization (RQ) for docID assignment in the main experiments, while alternative schemes, including BM25 based identifiers and product quantization (PQ), are examined in ablation studies. 
For baseline, the sparse retriever BM25 is implemented with the \emph{bm25s}\footnote{\url{https://github.com/xhluca/bm25s}} library, and the dense retriever DPR with the \emph{pyserini}\footnote{\url{https://github.com/castorini/pyserini}} toolkit. 
Other GR baselines are reproduced from their official codebases for fair comparison. 
All model editing methods operate on the same pre-trained GR model trained on the original corpus, restricting edits to the decoder layers identified by our analysis.


%% file: Tables/00-main.tex
\begin{table*}[t]
\caption{
Retrieval performance on the {\bf NQ}~\citep{kwiatkowski2019natural} and {\bf MS-MARCO}~\citep{nguyen2016ms} datasets.  
Each cell reports score for \emph{initial documents} / \emph{new documents}.  
Bold and underlined denote the best and second-best results, respectively, for each metric.  
“From Scratch” is reported only as a theoretical upper bound and is not considered in ranking due to its high retraining cost.  
* Significant improvements against the best-performing baseline for each dataset are marked with * (t-test, $p<0.05$).
}
\label{tab:main}
\centering
\begin{tabular}{lccccccc}
\toprule
\multicolumn{1}{c}{\bf Method} & 
\multicolumn{4}{c}{\bf NQ} & 
\multicolumn{2}{c}{\bf MS-MARCO} &  \\
\cmidrule(lr){2-5} \cmidrule(lr){6-7}
& R@1 & R@10 & R@100 & MRR@100 & R@10 & MRR@10 & \bf TPD (s) \\
\midrule
\multicolumn{8}{l}{\bf No Training} \\
BM25        & 0.374 \,/\, 0.365 & 0.761 \,/\, 0.770 & 0.906 \,/\, 0.860 & 0.476 \,/\, 0.471 & 0.691 \,/\, 0.641 & 0.486 \,/\, 0.471 & -- \\
DPR         & {0.651} \,/\, {0.647} & {0.874} \,/\, {0.871} & {0.926} \,/\, \underline{0.921} & {0.706} \,/\, {0.710} & {0.767} \,/\, {0.754} & {0.526} \,/\, {0.515} & -- \\
DSI         & 0.655 \,/\, 0.071 & 0.866 \,/\, 0.193 & 0.901 \,/\, 0.250 & 0.705 \,/\, 0.105 & 0.541 \,/\, 0.056 & 0.392 \,/\, 0.032 & -- \\
Ultron      & {0.671} \,/\, 0.064 & {0.867} \,/\, 0.134 & {0.911} \,/\, 0.208 & {0.722} \,/\, 0.168 & {0.731} \,/\, 0.069 & {0.454} \,/\, 0.030 & -- \\
\midrule
\multicolumn{8}{l}{\bf Incremental Training} \\
From Scratch & 0.696 \,/\, 0.693 & 0.886 \,/\, 0.883 & 0.931 \,/\, 0.930 & 0.744 \,/\, 0.742 & 0.775 \,/\, 0.772 & 0.532 \,/\, 0.529 & -- \\
New-Doc FT   & 0.696 \,/\, 0.660 & 0.886 \,/\, 0.848 & 0.931 \,/\, 0.917 & 0.744 \,/\, 0.732 & 0.775 \,/\, 0.751 & 0.532 \,/\, 0.511 & \underline{3.42} \\
DSI++        & 0.696 \,/\, \underline{0.674} & 0.886 \,/\, \underline{0.878} & 0.931 \,/\, {0.920} & 0.744 \,/\, \underline{0.735} & 0.775 \,/\, \underline{0.759} & 0.532 \,/\, \underline{0.517} & 3.54 \\
\midrule
\multicolumn{8}{l}{\bf Model Editing} \\
ROME         & 0.696 \,/\, 0.201 & 0.886 \,/\, 0.276 & 0.931 \,/\, 0.356 & 0.744 \,/\, 0.326 & 0.775 \,/\, 0.351 & 0.532 \,/\, 0.251 & 20.23 \\
MEMIT        & 0.696 \,/\, 0.609 & 0.886 \,/\, 0.687 & 0.931 \,/\, 0.756 & 0.744 \,/\, 0.665 & 0.775 \,/\, 0.673 & 0.532 \,/\, 0.495 & 12.62 \\
AlphaEdit    & 0.696 \,/\, 0.616 & {0.886} \,/\, 0.677 & 0.931 \,/\, 0.745 & 0.744 \,/\, 0.664 & 0.775 \,/\, 0.661 & 0.532 \,/\, 0.490 & 12.12 \\
DOME         & ~{0.696} \,/\, \textbf{0.686}\rlap{$^*$} & ~{0.886} \,/\, \textbf{0.880}\rlap{$^*$} & ~{0.931} \,/\, \textbf{0.927}\rlap{$^*$} & ~{0.744} \,/\, \textbf{0.740}\rlap{$^*$} & ~{0.775} \,/\, \textbf{0.764}\rlap{$^*$} & ~{0.532} \,/\, \textbf{0.524}$^*$ & \textbf{\phantom{0}2.14}\rlap{$^*$} \\
\bottomrule
\end{tabular}
\end{table*}

%% file: Sections/05-results.tex
\subsection{Results on new documents (RQ1)}
To answer RQ1, we evaluate the overall performance and adaptation efficiency of DOME.

\noindent
\textbf{Performance.} 
We first assess the effectiveness of DOME in adapting the GR model to newly added documents. The results are presented in Table~\ref{tab:main}. We observe that:
\begin{enumerate*}[nosep, leftmargin=*]
    \item Severe generalization failure without adaptation.  
    Models such as DSI and Ultron show drastic performance degradation on new documents, with Recall@1 on NQ dropping to 0.071 and 0.064, respectively.  
    These results clearly confirm the fundamental generalization bottleneck in GR models, as the decoder fails to generate docIDs unseen during initial training.  
    \item DOME demonstrates effective adaptation.  
    On NQ, DOME attains a Recall@1 of 0.686 on new documents, nearly matching full retraining (0.693) and outperforming the strongest incremental training baseline, DSI++ (0.674).  
    This result indicates that DOME effectively adapts GR models to new corpora while maintaining high retrieval accuracy.  
    \item DOME substantially surpasses existing model-editing methods. 
    Compared with MEMIT and AlphaEdit, DOME delivers notably higher retrieval accuracy.  
    Existing editing methods struggle to produce discriminative edit vectors when target docIDs overlap, whereas DOME's hybrid label adaptive training enables precise and stable docID mapping.  
\end{enumerate*}

\noindent\textbf{Adaptation efficiency.}
We further examine the efficiency aspect of RQ1 by measuring the update time per document (TPD). The results are presented in Table~\ref{tab:main}, and the main conclusions are as follows:
\begin{enumerate*}[nosep, leftmargin=*]
    \item DOME provides remarkably fast and scalable adaptation. 
    DOME achieves a TPD of 2.14 seconds, which is significantly faster than incremental training methods such as DSI++ (3.54s), confirming its high efficiency in updating GR models.  
    \item DOME achieves substantial speedup over editing baselines.  
    It is more than five times faster than MEMIT (12.62s) and AlphaEdit (12.12s), while maintaining superior adaptation performance.  
    The improved efficiency stems from our carefully designed framework and training strategy, which are tailored to enable effective adaptation of generative retrieval models to new documents.

\end{enumerate*}




\input{Tables/01-submain}

\subsection{Results on initial documents (RQ2)}
To address RQ2, we evaluate how well DOME preserves existing knowledge by measuring retrieval performance on the initial document corpus after adaptation. The results are shown in Table~\ref{tab:submain}, and the main findings are as follows:
\begin{enumerate*}[nosep, leftmargin=*]
    \item DOME effectively preserves prior knowledge.  
    Its performance on the initial corpus remains nearly unchanged, with Recall@1 decreasing only slightly from 0.696 to 0.692.  
    The forgetting score ($F_n$) of 0.003, the best among all methods, further demonstrates its strong knowledge retention capability.  
    \item Knowledge preservation results from constrained updates.  
    As defined in Eq.~\eqref{eq:model-edit-solution-new}, DOME introduces a constraint based on key-value pairs from the initial corpus ($K_0, V_0$).  
    The term $K_0 K_0^\top$ ensures that the final update matrix $\Delta$ minimally affects outputs for previously learned documents.  
    \item Other model-editing methods exhibit severe forgetting. 
    ROME collapses to a Recall@1 of 0.285, and both MEMIT and AlphaEdit perform notably worse than DOME.  
    These editing strategies produce edit vectors with poor discriminability across different queries, which leads to interference with existing knowledge and ultimately degrades overall model performance.
    \item DOME mitigates interference through hybrid-label adaptive training.  
    By generating query-specific update vectors, DOME avoids conflicting updates and achieves substantially better knowledge retention than competing approaches.  
\end{enumerate*}

\input{Tables/02-ablation}

\subsection{Ablation study (RQ3)}
To answer RQ3, we conduct an ablation study to analyze the contribution of DOME's key components: the number of pseudo queries, the docID assignment, the edited decoder layer range, and the hybrid-label training strategy. The results on the NQ new document set are shown in Table~\ref{tab:ablation}.

\noindent\textbf{Effect of pseudo query anount.}  
The number of pseudo queries per document strongly influences adaptation.  
Using a single query yields poor performance (R@1 = 0.506), while increasing to four and seven queries progressively improves it to 0.652.  
More queries provide richer contextual cues, enhancing the mapping from semantic representations to docIDs.  

\noindent\textbf{Effect of docID assignment.}  
DOME remains robust under different docID schemes.  
Replacing the default RQ-based identifiers with BM25- or PQ-based docIDs yields comparable performance, indicating that DOME's effectiveness does not rely on a specific docID structure and generalizes across GR systems.  

\noindent\textbf{Effect of edited layer range.}  
Performance varies with the decoder layers selected for editing.  
Editing layers 14–18 achieves the best results, while shifting to earlier (11–15) or later (18–22) layers causes moderate (R@1 = 659) and sharp (R@1 = 0.536) drops, respectively.  
This confirms that docID mapping knowledge is concentrated in the middle-to-late decoder layers, demonstrating the importance of our layer localization strategy.  

\begin{figure}[htbp]
    \centering
    \includegraphics[width=0.95\linewidth]{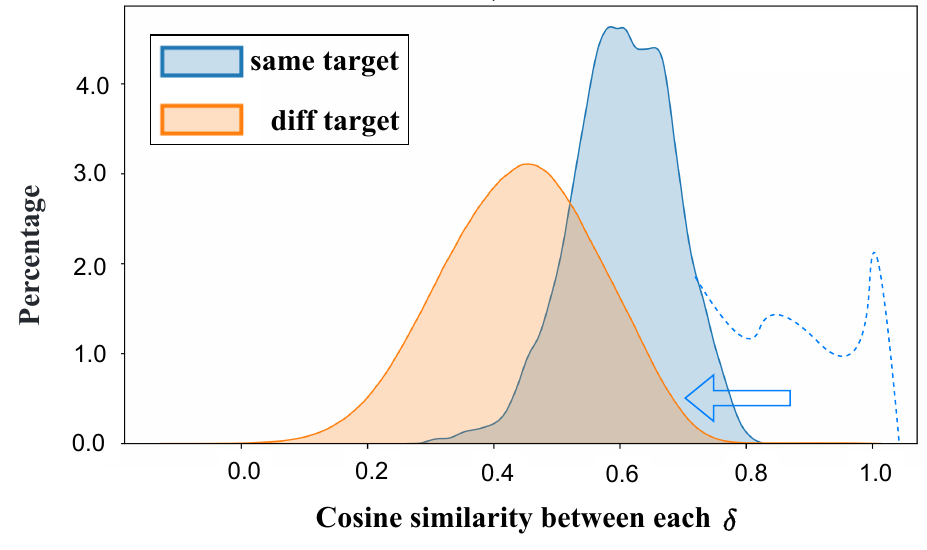} 
    \caption{Percentage of pairwise cosine similarity of edit vector $\delta$ across GR edit requests on NQ. Yellow denotes pairs with different target docID, blue denotes pairs with the same target docID.}
    \label{fig:softlabel}
\end{figure}

\noindent\textbf{Effect of hybrid-label training strategy.}  
The hybrid-label training is crucial for DOME's success, as removing either the soft or hard component significantly degrades performance.  
Without soft labels, one-hot supervision causes edit vectors for the same docID to collapse into similar directions.
This considerably reduces adaptation flexibility and degrades retrieval performance.
Without hard labels, the model lacks precise supervision, leading to severe degradation.  
As shown in Figure~\ref{fig:softlabel}, the hybrid strategy reduces excessive similarity among edit vectors sharing the same target docID, yielding more discriminative, query-specific updates that enhance adaptation and retrieval quality.




%% file: Tables/01-submain.tex
\begin{table}[htbp]
\centering
\caption{
Retrieval performance on NQ, reported separately for initial documents. 
Bold denotes the best result for each metric, and * indicates significant improvement (t-test, $p<0.05$). 
}
\label{tab:submain}
\begin{tabular}{lccccc}
\toprule
Method & R@1 & R@10 & R@100 & MRR@100 & $F_n \downarrow$\\
\midrule
Base Model      & 0.696 & 0.886 & 0.931 & 0.744 & -- \\
New-Doc FT   & 0.544 & 0.741 & 0.805 & 0.612 & 0.125 \\
DSI++        & 0.674 & 0.879 & 0.926 & 0.721 & 0.007 \\
ROME         & 0.285 & 0.569 & 0.774 & 0.355 & 0.317 \\
MEMIT        & 0.654 & 0.866 & 0.904 & 0.705 & 0.020 \\
AlphaEdit    & 0.647 & 0.852 & 0.891 & 0.696 & 0.034 \\
DOME         & \textbf{0.692}\rlap{$^*$} & \textbf{0.883}\rlap{$^*$} & \textbf{0.928}\rlap{$^*$} & \textbf{0.736}\rlap{$^*$} & \textbf{0.003}\rlap{$^*$}\\
\bottomrule
\end{tabular}
\end{table}

%% file: Tables/02-ablation.tex
\begin{table}[htbp]
\centering
\caption{Ablation study on NQ.
}
\label{tab:ablation}
\begin{tabular}{lcccc}
\toprule
\textbf{Variant} & \textbf{R@1} & \textbf{R@10} & \textbf{R@100} & \textbf{MRR@100} \\

\midrule
DOME (full)            & \textbf{0.686} & \textbf{0.880} & \textbf{0.927} & \textbf{0.740} \\

\midrule
\multicolumn{5}{l}{\emph{Number of pseudo queries per new doc}} \\
Pseudo = 1             & 0.506 & 0.760 & 0.875 & 0.592 \\
Pseudo = 4             & 0.615 & 0.832 & 0.898 & 0.678 \\
Pseudo = 7             & 0.652 & 0.866 & 0.918 & 0.698 \\

\midrule
\multicolumn{5}{l}{\emph{DocID type}} \\
BM25-based docIDs      & 0.683 & 0.860 & 0.921 & 0.722 \\
PQ-based docIDs        & 0.675 & 0.877 & 0.933 & 0.719 \\
\midrule
\multicolumn{5}{l}{\emph{Edited decoder layer range}} \\
Layers 11--15          & 0.659 & 0.857 & 0.904 & 0.704 \\
Layers 18--22          & 0.536 & 0.717 & 0.836 & 0.624 \\

\midrule
\multicolumn{5}{l}{\emph{soft-to-hard label strategy}} \\
w/o soft label   & 0.646 & 0.805 & 0.901 & 0.697 \\
w/o hard label   & 0.325 & 0.556 & 0.711 & 0.436 \\
\bottomrule
\end{tabular}
\end{table}

%% file: Sections/07-conclusion.tex
\section{Conclusion}
In this work, we have identified that the decoder's failure to learn precise docID mappings is the key obstacle in adapting generative retrieval models to new documents. 
To address this, we have introduced \textbf{DOME}, a GR-specific model-editing framework with a hybrid-label adaptive training strategy that produces discriminative and precise updates to critical decoder layers. 
Experiments on NQ and MS-MARCO have shown that DOME markedly improves retrieval on new documents, mitigates catastrophic forgetting, and reduces adaptation time by over 40\% versus incremental training, establishing it as an efficient and scalable solution. 

While promising, our current approach updates only the decoder's FFN layers. 
In future work, we will focus on further improving the efficiency of the editing process and on exploring adaptation strategies for additional modules, including attention layers and encoder representations. 
We expect that incorporating more components will improve model adaptability and enhance retrieval performance on both original and new document sets.

%% file: Sections/08-appendix.tex
\section{Appendix}

\subsection{Patching for locating critical layers}
\label{appdix:patching}

\begin{figure}[htbp]
    \centering
    \includegraphics[width=\linewidth]{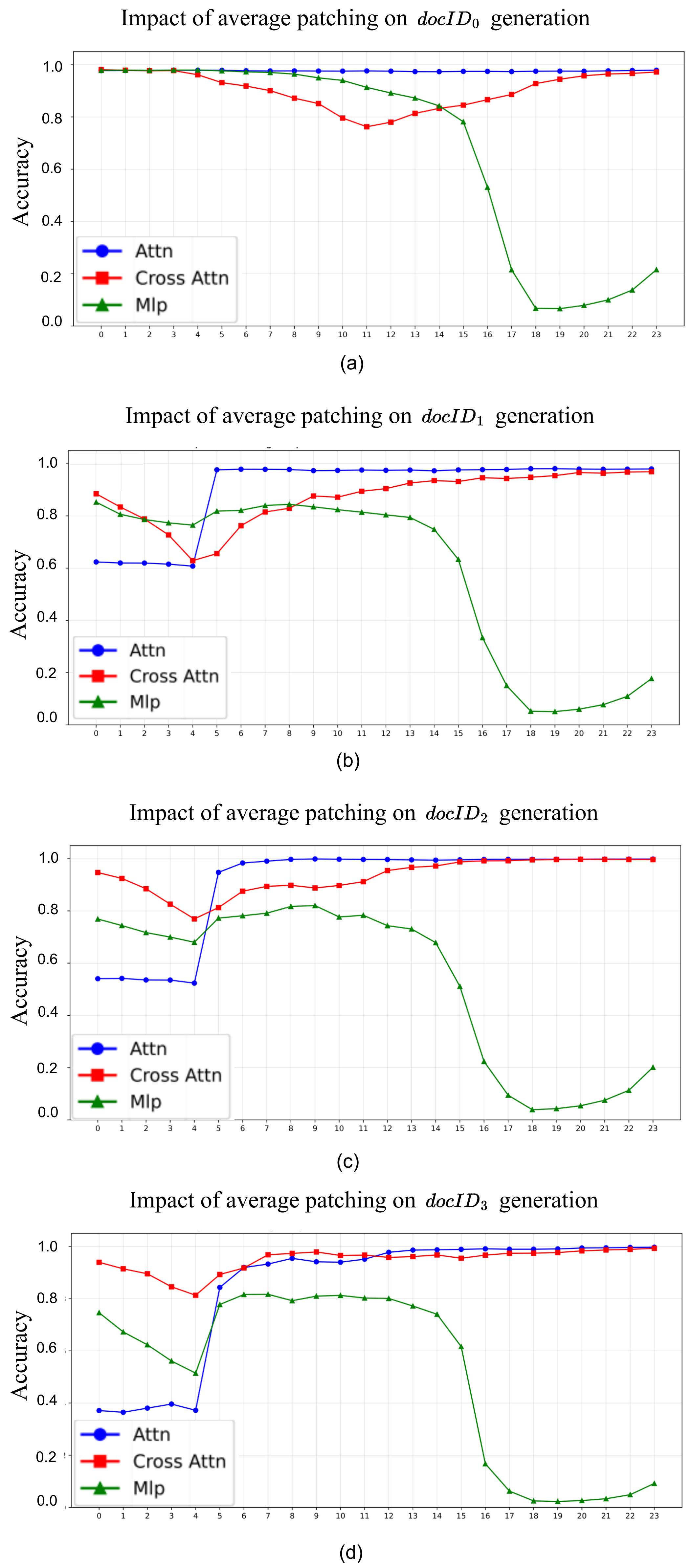}
    \caption{
    Accuracies of average patching across decoder layers for predicting docID tokens at positions 0–3 on NQ.
}
    \label{fig:appdix_patching}
\end{figure}

To investigate which decoder layers store the critical knowledge for docID mapping, 
we analyse the impact of replacing intermediate outputs in different modules (FFN, self-attention, cross-attention) 
during the prediction of each docID token in the generative process.
We evaluate on four separate settings, where the target token is located at positions $t=0$, $t=1$, $t=2$, and $t=3$ in the docID sequence.
The horizontal axis in Figure~\ref{fig:appdix_patching} denotes the decoder layer index, and the vertical axis denotes the probability of correctly predicting the target docID at that position.

\header{Patching technique}  
Following the average patching approach~\citep{reusch2025reverse}, for a specific decoder layer $l$ and module type $M$ (FFN / Self-Attn / Cross-Attn), 
we first compute the average output representation $\bar{h}_{l,M}$ over the training set.
Given an input (query,docID prefix) context $x$, we replace the original output $h_{l,M}(x)$ of module $M$ in layer $l$ with the average representation:
\begin{equation}
    h'_{l,M}(x) = \bar{h}_{l,M}.
\end{equation}
The patched representation propagates through the remaining layers, and we measure the probability
\begin{equation}
    P_{\text{correct}}(t, l, M) = p_\theta\left(y_t^\ast \mid q, y_{<t}; h_{l,M} \leftarrow \bar{h}_{l,M} \right),
\end{equation}
where $y_t^\ast$ denotes the ground‑truth docID token at position $t$.
By applying this operation to each layer and module in turn, we obtain the performance curves shown in Figure~\ref{fig:appdix_patching}.

\header{Observations}  
Across all four target positions ($t=0,1,2,3$), patching FFN layers in the middle and final stages of the decoder leads to the most pronounced drop in $P_{\text{correct}}$, indicating that these layers store critical mapping knowledge from semantic representations to docID tokens.
Patching self‑attention modules has little to no effect for $t=0$, while cross‑attention patching causes a moderate drop in mid layers.
For $t=1,2,3$, both self‑attention and cross‑attention patching result in a noticeable drop in the early decoder layers.
These consistent trends confirm the main finding in Section~\ref{section:3.3}: 
middle and final FFN modules are the most crucial for docID mapping and should be the primary targets for model editing.

\subsection{Generalization to varying numbers of documents (RQ4)}
\label{appdix:rq4}

\begin{figure}[htbp]
    \centering
    \includegraphics[width=\linewidth]{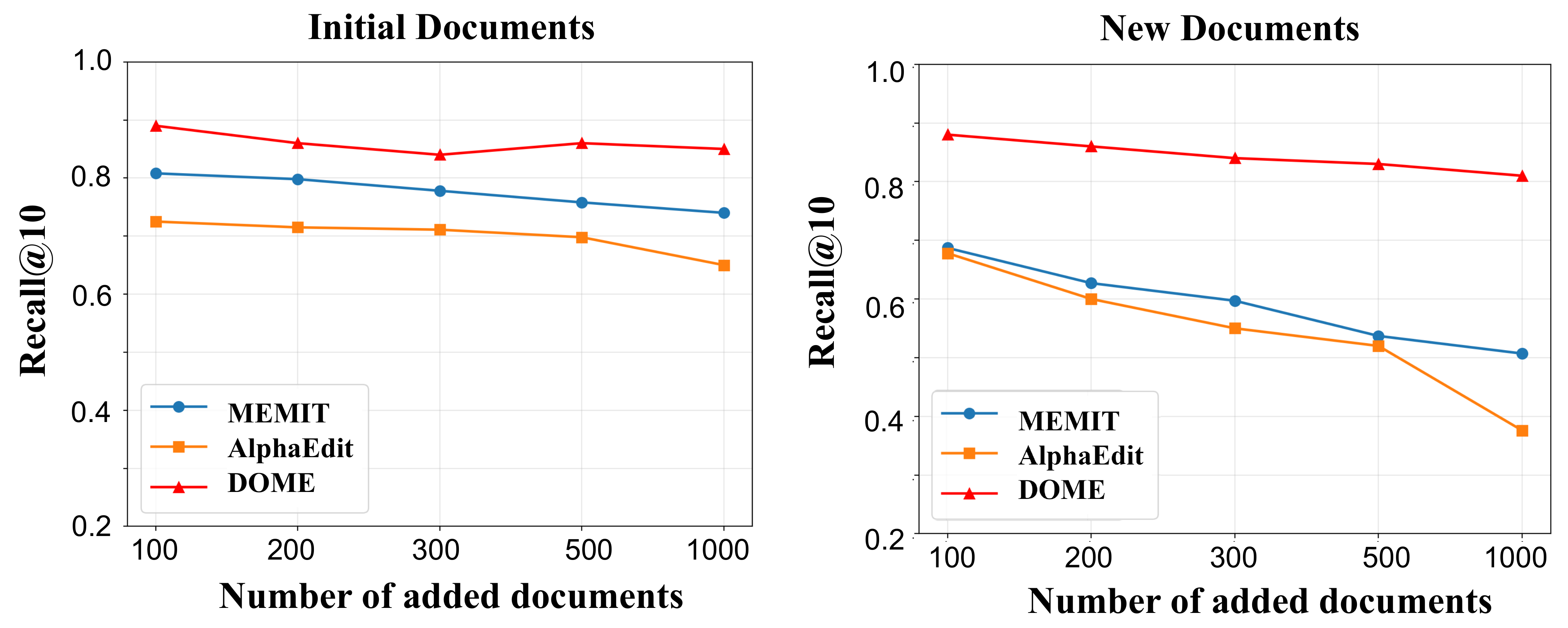}
    \caption{Performance over initial documents (left) and newly added documents (right) with a varying number of new documents on NQ.}
    \label{fig:scaling}
\end{figure}

To answer RQ4, we analyze DOME's generalization ability by varying the number of newly added documents from 100 to 1,000 and evaluating retrieval performance on both the initial and new collections. The results, illustrated in Figure~\ref{fig:scaling}, show that DOME maintains robust performance across different scales of document updates. We draw the following conclusions:
\begin{enumerate*}
    \item DOME effectively preserves existing knowledge.  
    As shown in the left panel of Figure~\ref{fig:scaling}, retrieval performance on the initial collection remains highly stable, exhibiting almost no degradation even when 1,000 new documents are integrated.  
    This demonstrates that the constrained update mechanism effectively prevents catastrophic forgetting regardless of update scale. 

    \item DOME demonstrates robust generalization to newly added documents.  
    As shown in the right panel of Figure~\ref{fig:scaling}, performance on new documents remains strong, with only a slight and expected decline as the editing task becomes more complex.  
    These results indicate that DOME can generalize effectively across varying numbers of new documents, integrating updates while ensuring that retrieval performance remains strong. 
\end{enumerate*}